\documentstyle[eqsecnum,preprint,prd,aps,psfig]{revtex}
\tighten
\def\jnfont{\rm}
\def\AP#1,{{\jnfont Ann.\ Phys.\ (N.Y.)} {\bf #1},}
\def\JETP#1,{{\jnfont Sov.\ Phys.\ JETP}\ {\bf #1},}
\def\JETPL#1,{{\jnfont JETP Lett.}\ {\bf #1},}
\def\NC#1,{{\jnfont Nuovo Cimento (Ser.~X)} {\bf #1},}
\def\NPB#1,{{\jnfont Nucl.\ Phys.}\ {\bf B#1},}
\def\PL#1,{{\jnfont Phys.\ Lett.}\ {\bf #1},}
\def\PLB#1,{{\jnfont Phys.\ Lett.\ B}~{\bf #1},}
\def\PR#1,{{\jnfont Phys.\ Rev.}\ {\bf #1},}
\def\PRD#1,{{\jnfont Phys.\ Rev.\ D}~{\bf #1},}
\def\PRL#1,{{\jnfont Phys.\ Rev.\ Lett.}\ {\bf #1},}
\def\PTP#1,{{\jnfont Prog.\ Theor.\ Phys.}\ {\bf #1},}
\def\SJNP#1,{{\jnfont Sov.\ J. Nucl.\ Phys.}\ {\bf #1},}
\def\ZPC#1,{{\jnfont Z. Phys.\ C} {\bf #1},}
\def\tfrac#1#2{{\textstyle{#1\over #2}}}
\def\Tr{\mathop{\rm Tr}}
\mathchardef\REAL="023C
\mathchardef\IMAG="023D
\def\Re{\mathop{\REAL\!{\it e}}} 
\def\Im{\mathop{\IMAG\!{\it m}}}
\def\fpi{f_\pi}
\def\etal{{\it et al.}}

\begin{document}

\preprint{\parbox{2.0in}{\noindent TU-552\\ RCNS-98-12\\ KNGU-INFO-PH-1\\
 July 1998\\}}

\title{\ \\[15mm] \Large\bf
Another Look at \boldmath$\pi\pi$\unboldmath\ Scattering\\[3mm] 
in the Scalar Channel\\[7mm]}

\author{{\large Keiji Igi}\\\ }

\address{{\it 
Department of Information Science, Kanagawa University\\
Hiratsuka, Kanagawa 259-1293, Japan\\}}

\author{{\large Ken-ichi Hikasa}\\\ }

\address{{\it Department of Physics, Tohoku University\\
      Aoba-ku, Sendai 980-8578, Japan}\\[5mm]\ }

\maketitle

\begin{abstract}
We set up a general framework to describe $\pi\pi$ scattering below 1 GeV 
based on chiral low-energy expansion with possible spin-0 and 1 resonances.  
Partial wave amplitudes are obtained with the $N/D$ method, which satisfy 
unitarity, analyticity and approximate crossing symmetry.  
Comparison with the phase shift data in the $J=0$ channel favors a 
scalar resonance near the $\rho$ mass.  
\end{abstract}


\section{Introduction}

Although Quantum Chromodynamics has long been accepted as the 
fundamental theory of the strong interaction, 
the spectrum of hadrons composed of light quarks still poses 
many unanswered questions.  Even below 1 GeV, the old 
controversy on the existence of the $\sigma$ meson, 
an isospin-0 scalar boson strongly coupled to the $\pi\pi$ 
system, remains unanswered.  
Recently, there have been some renewed interests in this 
problem both theoretically~\cite{Morgan,Dobado,Achasov,Zou,Kaminski,%
Janssen,Tornqvist,Anisovich,Svec,Harada,Ishida} and 
experimentally~\cite{Amsler,AldeA,AldeB}.  Some analyses favor 
the existence of $\sigma$.  The particle reappeared in the 1996 edition of 
``Review of Particle Physics'' (Particle Data Book)~\cite{PDG} 
as ``$f_0$(400--1200) or $\sigma$'' after an absence for more than 
two decades, though it is cautiously stated that 
``the interpretation of this entry as a particle is controversial.''  
There is no good agreement on its mass among the recent studies.  
For example, T\"ornqvist and Roos~\cite{Tornqvist} 
have used a ``unitarized quark model'' to 
fit the meson-meson $S$ wave amplitudes and claimed the existence 
of a very broad $\sigma$ with a mass of ${\sim}860$ MeV.  
Ishida {\it et al.}~\cite{Ishida} 
fit the $\pi\pi$ $S$ wave amplitudes to an $S$ matrix model and 
find the $\sigma$ mass of $585\pm20$ MeV.  
Although these results seem conclusive within their frameworks, 
the disagreement of the derived $\sigma$ mass may imply that 
quantitative conclusions are quite model dependent, 
casting some doubt in the very existence of $\sigma$.  
In any case, it is not easy to assess how model-independent are 
their conclusions.   

In this paper, we look at this problem from a somewhat different point 
of view.   We try to minimize the necessary model assumptions by a 
simple approach which only assumes the chiral flavor symmetry and general 
constraints on the amplitudes such as analyticity, unitarity, and crossing 
symmetry.   We notice that crossing symmetry is not taken into 
account in the recent works discussed above.  

As we put more emphasis on theoretical transparency than aiming at a 
perfect fit to the data, we concentrate on $\pi\pi$ scattering 
below 1 GeV and work in the chiral limit with massless pions.  Since 
the pions are the Goldstone bosons of the spontaneously broken 
$\rm SU(2)\times SU(2)$ symmetry, the form of the pion interactions is 
tightly constrained at low energies by the symmetry.  If we expand the 
$\pi\pi$ elastic scattering amplitude around $s=t=0$, chiral symmetry 
demands the amplitude vanishes at $s=t=0$ and the linear terms in $s$, $t$ 
are determined in terms of the pion decay constant $f_\pi$.   These and 
terms in higher order in $s$, $t$ can be described systematically if one 
uses the machinery of the chiral Lagrangian.

Although this expansion around the origin gives a good description of the 
amplitude at low energies, it breaks down when one approaches the mass of 
the lowest-lying hadrons (resonances) other than pions.  These resonances 
manifest themselves as a pole on the second sheet in the scattering 
amplitude.  We are thus led to start with a simple form of the amplitude 
which has relevant poles (corresponding to possible spin-0 isospin-0 
$\sigma$ and spin-1 isospin-1 $\rho$ resonances) and has the 
behavior consistent with chiral symmetry (low energy theorem).   

If there is a resonance in the $s$ channel, the same resonance is also 
exchanged in the $t$ and $u$ channels because of crossing symmetry.   
In a study of strongly interacting Higgs sector~\cite{SND} 
we found that the crossed-channel exchange of a vector resonance has 
a large reflection in the $J=0$ partial wave.   In most of the recent 
model studies of the $\sigma$ meson, this effect is not explicitly taken 
into consideration.  It is one of the motivations of this work to assess 
the importance of the crossed channel $\rho$ exchange in the scalar channel.  

As we are concerned with the strong interaction, the amplitude 
constructed in this way tend to violate unitarity near the pole.  
To obtain a unitary amplitude, we first project to partial waves (we will 
be concerned with $J=0$ and $J=1$ channels), and use the $N/D$ method 
to unitarize the partial wave amplitudes.  This method gives 
amplitudes which has the correct analytic properties with cuts on the 
real axis.   In this respect, it is superior to the $K$ matrix or Pad\'e 
unitarization scheme.  
Although the procedure is not exactly crossing symmetric, the deviation 
from the symmetry is controlled and mostly limited to the region near 
the pole.

In Section 2, we summarize the general characteristics of the $\pi\pi$ 
elastic scattering amplitude and set up our chirally symmetric 
`model-independent' amplitude with possible poles.  Two simple cases, 
`no $\sigma$' and `degenerate $\rho$--$\sigma$' are discussed in detail.  
In Section 3, we calculate the partial wave amplitudes from the invariant 
amplitudes in Section 2.  
Unitarization of the amplitudes is performed using the $N/D$ method.  
Relation of our method to the low-energy 
chiral expansion is clarified in Section 4.   
In Section 5, we determine the parameters in 
the amplitudes and compare them with the phase shift data.  We summarize 
and conclude with some remarks in Section 6.

\section{Characteristics of $\pi\pi$ amplitude}

The $\pi\pi$ system has three independent isospin channels.  
In terms of Mandelstam variables, the invariant amplitude for 
the process $\pi^i + \pi^j \to \pi^k + \pi^\ell$ has the form
\begin{equation} \label{eqM}
{\cal M}_{ijk\ell}(s,t) = A(s,t) \delta_{ij}\delta_{k\ell} 
+ A(t,s) \delta_{ik}\delta_{j\ell} + A(u,t) \delta_{i\ell}\delta_{jk} 
\;,
\end{equation}
where $i$, $j$, $\ldots = 1$, 2, 3 are isospin indices 
(with $\pi^\pm = (\pi^1 \pm i\pi^2)/\sqrt2$, $\pi^0=\pi^3$).  
The variable $s$ is the c.m.\ energy squared, $t=-s(1-\cos\theta)/2$ and 
$u=-s(1+\cos\theta)/2$ with $\cos\theta$ denoting the c.m.\ scattering angle.  
Note that $s+t+u=0$.  
There is only one analytic function $A(s,t)$ because of crossing 
symmetry.  It satisfies $A(s,t)=A(s,u)$ due to Bose symmetry.
The last term in (\ref{eqM}) thus may be rewritten as 
$A(u,s)\delta_{i\ell}\delta_{jk} $.  

Chiral symmetry low energy theorem demands that $A$ behaves near 
$s=t=0$ as
\begin{equation}
A(s,t) = {s\over \fpi^2} + {\cal O}(s^2,st,t^2) \;,
\end{equation}
where $\fpi\approx93$ MeV is the pion decay constant.  The structure of the 
second term will be discussed later.  

The expansion breaks down by the existence of a resonance.  
We expect that possible lowest-lying resonances are in $I=J=0$ and 
$I=J=1$ channels.  In the narrow width approximation, the contribution 
of these resonances may be written
\begin{equation}
A(s,t) = {g_\sigma^2 s\over m_\sigma^2-s} 
\end{equation}
for the scalar exchange (we write $s$ in the numerator instead of a 
constant, to be consistent with the low energy theorem.  This corresponds 
to adding a contact interaction like that in the $\sigma$ model)
and
\begin{equation}
A(s,t) = g_\rho^2 \biggl( {s-u\over m_\rho^2-t} + {s-t\over m_\rho^2-u}
\biggr)
\end{equation}
for the vector exchange (the numerator is the minimal dependence 
to assure spin 1 and has the same form as the gauge boson exchange).  

The tail of these exchange amplitude contributes to the slope of the 
amplitude at the origin.  If we assume that these two resonances saturate 
the low energy theorem, we find
\begin{equation} \label{eqsumlet}
{g_\sigma^2\over m_\sigma^2} + {3g_\rho^2\over m_\rho^2} = {1\over \fpi^2}\;.
\end{equation}
This condition, applied to the electroweak symmetry breaking, has been used 
in our previous study of the strong $WW$ scattering~\cite{SND}, 
in which we have obtained partial wave amplitudes consistent with unitarity 
and analyticity.  For hadron physics, 
it turns out that the condition is too strong to explain the observed width 
of the $\rho$ meson.  Even if one maximizes the vector coupling and includes 
the enhancing effect of unitarization, the resulting 
width is too small by $\sim 20\%$.   Thus we are led to relax the condition 
(\ref{eqsumlet}) to increase the $\rho\pi\pi$ coupling $g_\rho$.  
This may be done by subtracting the ${\cal O}(s)$ part from the exchange 
amplitudes and add a suitable ${\cal O}(s)$ term to $A(s,t)$ instead.  
This procedure gives
\begin{equation} \label{eqAP}
A(s,t) = {s\over \fpi^2} + {g_\sigma^2 s^2\over m_\sigma^2(m_\sigma^2-s)} 
+ {g_\rho^2\over m_\rho^2}\,\biggl( {t(s-u)\over m_\rho^2-t} 
+ {u(s-t)\over m_\rho^2-u}\biggr). 
\end{equation}
At the lowest order, the $\rho$ width may be reproduced if one takes 
$g_\rho$ around the KSRF value~\cite{KSRF} 
$g_\rho^2=m_\rho^2/2\fpi^2$.  
The price to pay is the worse high energy behavior.  

Expanding (\ref{eqAP}) to second order, we have
\begin{equation} \label{eqAPX}
A(s,t) \simeq {s\over \fpi^2} + 
{g_\sigma^2 s^2\over m_\sigma^4} 
+ {g_\rho^2\over m_\rho^4}\,( -2s^2+t^2+u^2) \;.
\end{equation}
This will be used later in matching with chiral Lagrangian.  

To assess the possible existence of $\sigma$, we will compare the two 
cases (1) no $\sigma$ ($\rho$ only) and 
(2) degenerate $\rho$--$\sigma$.  We now discuss motivations for these 
choices.  

(1) {\it No $\sigma$ meson:\/}  In the nonrelativistic quark 
model, the lowest-lying $S$ wave mesons are pseudoscalar ($\pi$, $\eta$) 
and vector ($\rho$, $\omega$).  Scalar mesons are $P$ wave states and are 
expected to have similar masses as the other $P$ wave states, the axial 
vector and tensor mesons which lie in the 1200--1300 MeV range.  As we will 
be concerned with the scattering amplitude below 1 GeV, such mesons in 
this mass range have small effect and we can simply 
take $g_\sigma=0$ to illustrate this case.  We may recall that the pion 
electromagnetic form factor is rather well described~\cite{Sakurai} by 
the hypothesis of $\rho$ dominance.  The coupling of $\rho$ to pions given 
by the KSRF relation~\cite{KSRF} 
\begin{equation} \label{eqKSRF}
g_\rho^2 = {m_\rho^2\over 2 \fpi^2} 
\end{equation}
reproduces the $\rho$ width quite well.

(2) {\it Degenerate $\rho$--$\sigma$:\/}  Since the light quarks 
are essentially massless compared to the QCD scale, there is no reason 
that nonrelativistic quark model reliably describe the spectrum.  
In the string-type 
picture of hadrons, the spectrum of the states has a tower structure 
and the vector meson is accompanied by a scalar daughter.  This situation 
in the narrow width approximation 
is realized in the Veneziano amplitude~\cite{Veneziano}. 

The degeneracy of $\rho$ and $\sigma$ is also suggested in the framework of 
nonlinear realization of the $\rm SU(2)\times\rm SU(2)$ chiral flavor 
symmetry developed by Weinberg~\cite{Weinberg}.  
Algebraization of the Adler-Weisberger sum rule results in 
the mass matrix structure with this degeneracy, again in the narrow width 
(large $N_{\rm color}$) limit.  The couplings $g_\rho$ and $g_\sigma$ 
are found to be equal and has the same strength as the KSRF coupling
\begin{equation} \label{eqWeinberg}
g_\sigma^2=g_\rho^2 = {m_\rho^2\over 2 \fpi^2} \;.
\end{equation}

The Veneziano amplitude also gives $g_\rho=g_\sigma$ but the size of 
the coupling is different, as we will now discuss.  
The Veneziano $\pi\pi$ scattering amplitude takes a simpler form 
for the charge eigenstates $\pi^+\pi^-\to \pi^+\pi^-$.  
With the constraints of chiral symmetry, it reads~\cite{Lovelace}
\begin{equation} \label{eqBfour}
B_4(s,t)=-{2m_\rho^2\over \pi \fpi^2}\,
{\Gamma((1-s/m_\rho^2)/2)\Gamma((1-t/m_\rho^2)/2)\over
\Gamma(u/2m_\rho^2)}\;.
\end{equation}
Vanishing of the amplitude at $s=t=0$ demands that the intercept of the 
Regge trajectory is $1/2$, and the overall coefficient is determined 
by the scale of chiral symmetry breaking $\fpi$.  
The invariant amplitude $A$ is related to (\ref{eqBfour}) by the relation 
$A(s,t) = [ B_4(s,t) + B_4(s,u) - B_4(t,u) ]/2$.

As is well known, the amplitude (\ref{eqBfour}) has an infinite number of 
poles both 
in the $s$ and $t$ channels.  The lowest-lying poles are at $s=m_\rho^2$ 
and $t=m_\rho^2$, at which the amplitude behaves as
\begin{equation} \label{eqBfourpole}
B_4(s,t)\sim \cases{\displaystyle
{2m_\rho^2\over \pi \fpi^2}\,{m_\rho^2+t\over m_\rho^2-s}
& ($s\sim m_\rho^2$),\cr\noalign{\vskip 3\jot}
\displaystyle {2m_\rho^2\over \pi \fpi^2}\,{m_\rho^2+s\over m_\rho^2-t}
& ($t\sim m_\rho^2$).\cr}
\end{equation}
Expanding (\ref{eqBfourpole}) in partial waves, one finds that a scalar and 
a vector state are degenerate at $m_\rho$.  The corresponding couplings are 
\begin{equation} \label{eqVeneziano}
g_\sigma^2=g_\rho^2 = {m_\rho^2\over\pi \fpi^2} \;.
\end{equation}
The chiral Veneziano amplitude may be approximated by the form (\ref{eqAP}) 
with appropriate couplings and masses in the energy region of our 
interest, where the higher poles have small effect.

\section{Partial wave amplitudes}

The invariant amplitude can be expanded in terms of partial waves 
for states having definite isospin $I$:  
\begin{equation} 
a_{IJ}(s) = {1\over 64\pi}\int_{-1}^1 \!\! d\cos\theta\,
P_J(\cos\theta) \, {\cal M}^{(I)}(s,t) \;,
\end{equation}
with $t=-s(1-\cos\theta)/2$ and 
\begin{mathletters}
\begin{eqnarray}
&&{\cal M}^{(I=0)} = 3A(s,t) + A(t,s) + A(u,s)\;,\\
&&{\cal M}^{(I=1)} = A(t,s) - A(u,s)\;,\\
&&{\cal M}^{(I=2)} = A(t,s) + A(u,s)\;.
\end{eqnarray}
\end{mathletters}
Elastic unitarity requires $\Im a_{IJ}^{-1}(s) = -1$ and the amplitude 
can be written in terms of the phase shift $\delta_{IJ}$ as
\begin{equation} 
a_{IJ}=e^{i\delta_{IJ}}\sin\delta_{IJ} \;.
\end{equation}
Inelastic channels ($4\pi$, $\ldots$) are known to be negligible below the 
$K\bar K$ threshold~\cite{inel,Protopopescu}, in accordance with 
the expectation based on chiral symmetry (the $\pi\pi\to 4\pi$ cross 
section starts at the order ${\sim}s^4/(4\pi\fpi)^8$).  

We project the subtracted pole amplitude (\ref{eqAP}) into partial waves, 
which we denote by $a^\circ_{IJ}$.  We find
\begin{mathletters} \label{eqaIJP}
\begin{eqnarray}
&& a^\circ_{00} = {1\over 16\pi} \biggl[ {s\over \fpi^2} + g_\sigma^2 \Bigl(
\tfrac32 f_r(s/m_\sigma^2) + f_{0\sigma}(s/m_\sigma^2) \Bigr) 
+ 2 g_\rho^2 f_{0\rho}(s/m_\rho^2) \biggr],\\
&& a^\circ_{20} = {1\over 16\pi} \biggl[ -{s\over 2\fpi^2} 
+ g_\sigma^2 f_{0\sigma}(s/m_\sigma^2) 
- g_\rho^2 f_{0\rho}(s/m_\rho^2) \biggr],\\
&& a^\circ_{11} = {1\over 16\pi} \biggl[ {s\over 6\fpi^2} 
+ g_\sigma^2 f_{1\sigma}(s/m_\sigma^2) 
+ g_\rho^2 \Bigl( \tfrac13 f_r(s/m_\rho^2) + f_{1\rho}(s/m_\rho^2) \Bigr)
\biggr],
\end{eqnarray}
\end{mathletters}
where
\begin{mathletters}
\begin{eqnarray}
&& f_r(x) = {x^2\over 1-x}, \\
&& f_{0\sigma}(x) = {1\over x}\log(1+x) -1 + {x\over 2}, \\
&& f_{0\rho}(x) = \biggl({1\over x}+2\biggr)\log(1+x) - 1 - {3\over2}\,x, \\
&& f_{1\sigma}(x) = {1\over x}\biggl({2\over x}+1\biggr)\log(1+x) 
- {2\over x} -{x\over 6} ,\\
&& f_{1\rho}(x) = \biggl({1\over x}+2\biggr)\biggl({2\over x}+1\biggr) 
\log(1+x) -{2\over x} - 4 -{x\over6} .
\end{eqnarray}
\end{mathletters}
These functions may alternatively be obtained from the $n_\alpha$ 
functions defined in~\cite{SND}, Eq.~(13) by subtracting 
the ${\cal O}(x)$ term.  Near $x=0$, these functions behave as
\begin{mathletters}
\begin{eqnarray}
&& f_r(x) \simeq x^2,\\
&& f_{0\sigma}(x) \simeq {1\over 3}\, x^2,\\
&& f_{0\rho}(x) \simeq -{2\over 3}\, x^2,\\
&& f_{1\sigma}(x) \simeq -{1\over 6}\, x^2,\\
&& f_{1\rho}(x) \simeq {1\over 6}\, x^2.
\end{eqnarray}
\end{mathletters}

For large values of the couplings, these amplitudes badly violates 
unitarity near the resonances.  We use the $N/D$ method to obtain amplitudes 
satisfying elastic unitarity and analyticity.  This method is superior to $K$ 
matrix or Pad\'e unitarization scheme in that it automatically provides an 
amplitude having correct analytic behavior.  We thus write 
\begin{equation} \label{eqaIJU}
a_{IJ} = {N_{IJ}\over D_{IJ}},
\end{equation}
and use single $N/D$ iteration by setting $N_{IJ}=a^\circ_{IJ}$ 
(given by (\ref{eqaIJP})).  The denominator function is 
determined by analyticity
\begin{equation} 
\Im D_{IJ}(s) = - N_{IJ}(s) \theta(s) 
\end{equation}
(we assume the contribution of inelastic channels is not important), 
which symbolically gives
\begin{equation}
D_{IJ}(s) = - {1\over\pi}\int_0^\infty\,{ds'\over s'-s}\,N_{IJ}(s').
\end{equation}
Since $N_{IJ}(s) \sim s$ at $s\to\infty$, the dispersion integral 
has to be subtracted twice.  One of the subraction constants is fixed 
by the normalization condition $D_{IJ}(0)=1$ (remember that our amplitude 
$a^\circ_{IJ}$ is constructed to be exact near $s=0$, which requires this 
condition), and the second constant determines the ${\cal O}(s)$ behavior 
of $D_{IJ}(s)$ as will be discussed later.   

To write down the explicit functional form of $D$, 
we define the function $d_\alpha(x)$ with the property
\begin{equation} 
\mathop{\rm disc} d_\alpha(x) 
\equiv d_\alpha(x+i\epsilon) - d_\alpha(x-i\epsilon) 
= 2\pi i f_\alpha(x) \theta(x) 
\end{equation}
by
\begin{equation} \label{eqddef}
d_\alpha(x) + (c_\alpha \log R + c'_\alpha) x = 
x\int_0^R {dy\over y(y-x)}\,f_\alpha(y) \qquad (R\to\infty),
\end{equation}
where we demand $d_\alpha(x)\sim x^2$ near $x=0$.  This is accomplished 
by separating the ${\cal O}(x)$ term of the integral as the second term 
in the LHS of (\ref{eqddef}).  Though the integral diverges logarithmically 
for $R\to\infty$, $d_\alpha(x)$ thus defined is finite in this limit.  
We find 
\begin{mathletters}
\begin{eqnarray}
&& d_r(x) = -{x^2\over 1-x}\log(-x), \\
&& d_{0\sigma}(x) = {1\over x} L(x) + \biggl(1-{x\over 2}\biggr)\log(-x) -1 
+{x\over 4}, \\
&& d_{0\rho}(x) = \biggl({1\over x}+2\biggr) L(x) 
+ \biggl(1+{3\over2}x\biggr) \log(-x) - 1 - {7\over 4} x,\\
&& d_{1\sigma}(x) = {1\over x}\biggl({2\over x}+1\biggr) L(x) 
+ \biggl( {2\over x} + {x\over 6} \biggr) \log(-x) 
-{2\over x} - {1\over 2} + {x\over 36},\\
&& d_{1\rho}(x) = \biggl({1\over x}+2\biggr)\biggl({2\over x}+1\biggr) L(x) 
+ \biggl({2\over x}+4+{x\over 6}\biggr) \log(-x) -{2\over x} - {9\over 2}
- {35\over 36} x,
\end{eqnarray}
\end{mathletters}
with
\begin{equation}
L(x) = -{\rm Li}_2(-x) - \log(-x)\log(1+x), 
\end{equation}
and
\begin{eqnarray}
 &&c_r = -1,\qquad 
 c_{0\sigma} = {1\over 2}, \qquad
 c_{0\rho} = - {3\over2},\qquad
 c_{1\sigma} = c_{1\rho } = -{1\over 6},\\
 &&c'_r = 0,\qquad 
 c'_{0\sigma} = -{1\over 4}, \qquad
 c'_{0\rho} = {7\over 4},\qquad
 c'_{1\sigma} = -{1\over 36},\qquad
 c'_{1\rho } = {35\over 36}.
\end{eqnarray}

Then we can write
\begin{mathletters}
\begin{eqnarray}
&&N_{00} = {1\over 16\pi} \biggl[ {s\over \fpi^2} + g_\sigma^2 \Bigl(
\tfrac32 f_r(s/m_\sigma^2) + f_{0\sigma}(s/m_\sigma^2) \Bigr) 
+ 2 g_\rho^2 f_{0\rho}(s/m_\rho^2) \biggr],\\
&&N_{20} = {1\over 16\pi} \biggl[ -{s\over 2\fpi^2} 
+ g_\sigma^2 f_{0\sigma}(s/m_\sigma^2) 
- g_\rho^2 f_{0\rho}(s/m_\rho^2) \biggr],\\
&&N_{11} = {1\over 16\pi} \biggl[ {s\over 6\fpi^2} 
+ g_\sigma^2 f_{1\sigma}(s/m_\sigma^2) 
+ g_\rho^2 \Bigl( \tfrac13 f_r(s/m_\rho^2) + f_{1\rho}(s/m_\rho^2) \Bigr)
\biggr],
\end{eqnarray}
\end{mathletters}
\begin{mathletters}
\begin{eqnarray}
&&D_{00} = 1 - d_{00}'s 
- {1\over 16\pi^2} \biggl[ -{s\over \fpi^2}\log\Bigl({-s\over\mu^2}\Bigr)  
\nonumber\\ 
&&\qquad\quad {}
+ g_\sigma^2 
\Bigl(\tfrac32 d_r(s/m_\sigma^2) + d_{0\sigma}(s/m_\sigma^2) \Bigr) 
+ 2 g_\rho^2 d_{0\rho}(s/m_\rho^2) \biggr], \\
&&D_{20} = 1 - d_{20}'s 
+ {1\over 16\pi^2} \biggl[ -{s\over 2\fpi^2} \log\Bigl({-s\over\mu^2}\Bigr)
+ g_\sigma^2 d_{0\sigma}(s/m_\sigma^2) 
- g_\rho^2 d_{0\rho}(s/m_\rho^2) \biggr],\\
&&D_{11} = 1 - d_{11}' s 
- {1\over 16\pi^2} \biggl[ -{s\over 6\fpi^2} \log\Bigl({-s\over\mu^2}\Bigr)
\nonumber\\ 
&&\qquad\quad {}
+ g_\sigma^2 d_{1\sigma}(s/m_\sigma^2) 
+ g_\rho^2 \Bigl( \tfrac13 d_r(s/m_\rho^2) + d_{1\rho}(s/m_\rho^2) \Bigr)
\biggr].
\end{eqnarray}
\end{mathletters}
The coefficient $d_{IJ}'$ corresponds to the second subtraction constant 
and depends implicitly on $\mu$, which cancels the explicit 
$\mu$ dependence of the amplitudes.

The $N/D$ unitarization breaks crossing symmetry because it treats 
the $s$ channel distinctly from the other channels.  The deviation from 
symmetry is proportional to $D-1$, because the $N$ function is crossing 
symmetric by construction.  
Thus our unitarized amplitude is approximately crossing symmetric 
away from the resonance.

Our procedure apparently gives three independent subtraction constants.  
However, there can be at most two independent ones.  To see this, 
we now turn to the discussion of $\pi\pi$ scattering in the chiral 
Lagrangian language.

\section{Chiral Lagrangian up to $\partial^4$ order}

Interactions of pions at low energies can be described by the chiral 
Lagrangian, an effective Lagrangian with nonlinearly realized 
chiral symmetry, which is an expansion in the number of derivatives.  
The Lagrangian with terms up to the order $\partial^4$ takes the form 
(in the exact $\rm SU(2)\times SU(2)$ limit we are working)
\begin{equation} \label{eqLchi}
{\cal L} = {\cal L}_2 + {\cal L}_4 ,
\end{equation}
\begin{mathletters}
\begin{eqnarray}
&&{\cal L}_2  = 
{\fpi^2\over4}\Tr\bigl( \partial_\mu U^\dagger \partial^\mu U\bigr),\\
&&{\cal L}_4 = L_1 
\Bigl[\Tr\bigl(\partial_\mu U^\dagger \partial^\mu U\bigr)\Bigr]^2 
+ L_2 \Tr\bigl(\partial_\mu U^\dagger \partial_\nu U\bigr)
\Tr\bigl(\partial^\mu U^\dagger \partial^\nu U\bigr),
\end{eqnarray}
\end{mathletters}
with
\begin{equation} 
U = \exp(i\pi^i \tau^i/\fpi), 
\end{equation}
where $\pi^i$ ($i=1$, 2, 3) denotes the pion field and $\tau^i$ is the Pauli 
matrix.  The parameters in (\ref{eqLchi}) are in principle calculable from 
QCD, but in practice can be regarded as parameters to be determined from 
experiments.

The tree level $\pi\pi$ scattering amplitude derived from the Lagrangian 
(\ref{eqLchi}) is
\begin{equation} \label{eqALTr}
A(s,t) = {s\over \fpi^2} 
+ {8s^2\over \fpi^4}\,L_1 + {4(t^2+u^2)\over \fpi^4}\,L_2 \;.
\end{equation}
Comparing the tree chiral amplitude (\ref{eqALTr}) with the subtracted 
pole amplitude (\ref{eqAP}), we identify
\begin{equation} \label{eqLLB}
L_1 = {g_\sigma^2 \fpi^4 \over 8m_\sigma^4} 
- {g_\rho^2 \fpi^4\over 4m_\rho^4}, \qquad
L_2 = {g_\rho^2\fpi^4\over 4m_\rho^4}.
\end{equation}
It may be seen that the two coefficients reflect the underlying 
dynamics.  The scalar exchange gives $L_2=0$, and the vector exchange 
is characterized by the relation $L_1+L_2=0$.  In the $\rho$--$\sigma$ 
degenerate case with equal couplings, we have $2L_1+L_2=0$.

Independent determination of these parameters have been done 
using the $D$ wave $\pi\pi$ phase shift~\cite{Gasser} or 
$K\to\pi\pi\ell\nu$ decays~\cite{Bijnens}.   
These data exclude the case $L_2=0$.  The other two cases of 
$\rho$ only and degenerate $\rho$--$\sigma$ are compatible with 
the data.

At ${\cal O}(s^2)$, the contribution of one-loop 
graphs with the ${\cal L}_2$ vertices has to be included: 
\begin{eqnarray} &&
A(s,t) = {1\over 16\pi^2 \fpi^4}\biggl\{ {1\over2}s^2 
\biggl[{1\over\epsilon} -\log\Bigl({-s\over\mu^2}\Bigr)\biggr] 
+ {1\over6}t(t-u) 
\biggl[{1\over\epsilon} -\log\Bigl({-t\over\mu^2}\Bigr)\biggr]
\nonumber\\ && \qquad\qquad {} 
+ {1\over6}u(u-t) 
\biggl[{1\over\epsilon} -\log\Bigl({-u\over\mu^2}\Bigr)\biggr] 
 + {5\over9}s^2 + {13\over18}(t^2+u^2) \biggr\}. 
\end{eqnarray}
We have used dimensional regularization with $D=4-2\epsilon$ spacetime 
dimensions.  (In the usual convention, $1/\epsilon$ should be interpreted 
as $1/\epsilon-\gamma_E+\ln4\pi$.)  
Notice that the one-loop amplitude contains terms required by unitarity 
and analyticity at ${\cal O}(s^2)$.  
The logarithmic divergences can be absorbed into the parameter $L_i$ as
\begin{mathletters}
\begin{eqnarray}
&& L_1^r(\mu) = L_1 + {1\over 16\pi^2}\,{1\over 24}
\biggl({1\over\epsilon}+1\biggr),\\
&& L_2^r(\mu) = L_2 + {1\over 16\pi^2}\,{1\over 12}
\biggl({1\over\epsilon}+1\biggr),
\end{eqnarray}
\end{mathletters}
where we have followed the renormalization prescription of Gasser and 
Leutwyler~\cite{Gasser}.  
The amplitude in terms of the renormalized parameters is
\begin{eqnarray} 
&& A(s,t) = {s\over \fpi^2} 
+ {8s^2\over \fpi^4}\,L_1^r + {4(t^2+u^2)\over \fpi^4}\,L_2^r 
\nonumber\\ &&\qquad\qquad
+ {1\over 16\pi^2 \fpi^4}\biggl[
-{1\over2}s^2 \log\Bigl({-s\over\mu^2}\Bigr)
-{1\over6}t(t-u) \log\Bigl({-t\over\mu^2}\Bigr) 
-{1\over6}u(u-t) \log\Bigl({-u\over\mu^2}\Bigr) 
\nonumber\\ &&\qquad\qquad\qquad\qquad
+ {2\over9}s^2 + {7\over18}(t^2+u^2)\biggr].
\label{eqAL}
\end{eqnarray}
This gives the general form of the amplitude up to order ${\cal O}(s^2)$ 
compatible with chiral symmetry. 
Expanding the ${\cal O}(E^4)$ chiral amplitude (\ref{eqAL}) into partial 
waves, we find for $J\leq1$

\begin{mathletters} \label{eqaIJL}
\begin{eqnarray}
&& a_{00} = {1\over 16\pi} \biggl\{ {s\over \fpi^2} + {s^2\over \fpi^4}
\biggl[ {44\over 3} L_1^r + {28\over 3} L_2^r 
+ {1\over 16\pi^2}\Bigl( -\log{-s\over\mu^2} 
- {7\over 18} \log{s\over\mu^2} 
+{17\over 12}\Bigr)\biggr]\biggr\} ,\\
&& a_{20} = {1\over 32\pi} \biggl\{ -{s\over \fpi^2} 
+ {s^2\over \fpi^4} \biggl[ {16\over 3} L_1^r + {32\over 3} L_2^r 
+ {1\over 16\pi^2}\Bigl( -{1\over 2}\log{-s\over\mu^2} 
- {11\over 18} \log{s\over\mu^2} 
+{17\over 12}\Bigr)\biggr]\biggr\},\\
&& a_{11} = {1\over 96\pi} \biggl\{ {s\over \fpi^2} 
+ {s^2\over \fpi^4} \biggl[ -8 L_1^r + 4 L_2^r 
+ {1\over 16\pi^2}\Bigl( -{1\over 6}\log{-s\over\mu^2} 
+ {1\over 6} \log{s\over\mu^2} 
+{1\over 9}\Bigr)\biggr]\biggr\}.
\end{eqnarray}
\end{mathletters}

Now we are ready to discuss the connection with the partial waves obtained 
in Sec.~3.  
Since chiral symmetry allows only two independent ${\cal O}(s^2)$ parameters, 
the three coefficients $d_{00}'$, $d_{20}'$, and $d_{11}'$ cannot be 
arbitrary.  Expanding $a=N/D$ obtained in the previous section 
up to ${\cal O}(s^2)$, we have
\begin{mathletters} \label{eqaNDx}
\begin{eqnarray}
&& a_{00} \simeq {1\over 16\pi} \biggl\{ {s\over \fpi^2} 
+ s^2\,\biggl[ {d_{00}'\over \fpi^2} + {11g_\sigma^2\over 6m_\sigma^4} 
- {4g_\rho^2\over 3m_\rho^4}  
- {1\over 16\pi^2\fpi^4} \log\Bigl({-s\over\mu^2}\Bigr) \biggr]\biggr\},\\
&& a_{20} \simeq {1\over 32\pi} \biggl\{ -{s\over \fpi^2} 
+ s^2\,\biggl[ {d_{20}'\over \fpi^2} + {2g_\sigma^2\over 3m_\sigma^4} 
+ {4g_\rho^2\over 3 m_\rho^4}  
- {1\over 32\pi^2\fpi^4} \log\Bigl({-s\over\mu^2}\Bigr) \biggr]\biggr\},\\
&& a_{11} \simeq {1\over 96\pi} \biggl\{ {s\over \fpi^2} 
+ s^2\,\biggl[ {d_{11}'\over \fpi^2} - {g_\sigma^2\over m_\sigma^4} 
+ {3g_\rho^2\over m_\rho^4}  
- {1\over 96\pi^2\fpi^4} \log\Bigl({-s\over\mu^2}\Bigr) \biggr]\biggr\}.
\end{eqnarray}
\end{mathletters}
Comparing (\ref{eqaNDx}) with (\ref{eqaIJL}), we immediately find that the 
$\log(-s)$ 
terms obtained here are just as given by the general chiral Lagrangian, 
although $\log s$ terms are absent in (\ref{eqaNDx}).  The appearance of the 
former terms is the result of $s$ channel unitarity and analyticity of 
the $N/D$ amplitudes.  The latter terms, which reflects the crossed 
channel singularity, are not incorporated in our procedure which is not 
exactly crossing symmetric.  The effect of these logarithmic terms is 
unimportant if we choose $\mu$ to be around $m_\rho$, since the 
coefficient is small.  
Neglecting the logarithmic and related constant terms, we may identify
\begin{mathletters} \label{eqpolevschiral}
\begin{eqnarray}
&& {3\over 4} d_{00}'\fpi^2 + {11 g_\sigma^2 \fpi^4\over 8 m_\sigma^4} 
- {g_\rho^2 \fpi^4\over m_\rho^4} = 11 L_1^r + 7 L_2^r ,\\
&& {3\over 4} d_{20}'\fpi^2 + {g_\sigma^2 \fpi^4\over 2 m_\sigma^4} 
+ {g_\rho^2 \fpi^4\over m_\rho^4} = 4 L_1^r + 8 L_2^r ,\\
\label{eqEPwave}
&&  d_{11}'\fpi^2 - {g_\sigma^2 \fpi^4\over m_\sigma^4} 
+ {3g_\rho^2 \fpi^4\over m_\rho^4} = -8 L_1^r + 4 L_2^r .
\end{eqnarray}
\end{mathletters}
This gives one consistency condition for the three subtraction coefficients
\begin{equation} 
5d_{20}' = 4 \bigl(d_{00}'+d_{11}'\bigr) ,
\end{equation} 
which has to hold regardless of the dynamics.  
In addition, we can impose the dynamics-dependent relation between 
the chiral Lagrangian 
parameters discussed below (\ref{eqLLB}) on $L_1^r$ and $L_2^r$ 
in (\ref{eqpolevschiral}).  We find
\begin{equation}
{1\over 11} d_{00}' = -{1\over6} d_{11}' = {1\over4} d_{20}'
\end{equation}
for scalar only ($g_\rho=0$), 
\begin{equation} \label{eqdsvec}
{1\over 4} d_{00}' = -{1\over9} d_{11}' = -{1\over4} d_{20}'
\end{equation}
for vector only ($g_\sigma=0$), and
\begin{equation} \label{eqdsequal}
4 d_{00}' =  d_{11}' = d_{20}'
\end{equation}
for the equal contribution of both ($g_\sigma/m_\sigma^2=g_\rho/m_\rho^2$).
These conditions reduce the number of independent subtraction constants 
to one.

\section{Comparison with data}

Let us first discuss the $P$ wave amplitude $a_{11}$.  Experimentally, 
this amplitude is dominated by the $\rho$ resonance.  Since the 
existence of $\rho$ is well established and the parameters are well 
measured, we use the mass $m_\rho=769$ MeV and width 
$151$ MeV as inputs (as well as $f_\pi=93$ MeV).  Since 
we work in the chiral limit, we correct the measured $\rho$ width for the 
$P$ wave phase space factor $\beta^3$ to obtain the ideal width 
$\Gamma_\rho=187$ MeV.  We find that the result for the $S$ wave is not 
sensitive to the inclusion of this correction.  

The subtraction constant $d_{11}'$ may be fixed for a given set of 
model parameters ($g_\rho$, $g_\sigma$) by the condition that the 
unitarized amplitude gives the correct width $\Gamma_\rho$.  For a 
unitarized amplitude $a$, we define the width by 
\begin{equation} 
{d\over ds}\,a^{-1}(s)\Big|_{s=m^2} = -{1\over m\Gamma} \;,
\end{equation}
where the mass $m$ is defined by $a(m^2)=i$.  
This gives for the $N/D$ amplitude 
\begin{equation} \label{eqGrho}
\Gamma_\rho = {\Gamma_\rho^0\over \Re D_{11}(m_\rho^2)}\;,
\end{equation}
where
\begin{equation}
\Gamma_\rho^0 = {g_\rho^2 m_\rho\over 48\pi} \;.
\end{equation}
We thus obtain
\begin{equation} \label{eqdpr}
d'_{11} m_\rho^2 = 1 - {g_\rho^2 m_\rho\over 48\pi\Gamma_\rho} 
+ {m_\rho^2\over 96\pi^2 \fpi^2}\log{m_\rho^2\over \mu^2} 
-{1\over 16\pi^2}\biggl[g_\sigma^2 
\Re d_{1\sigma}(m_\rho^2/m_\sigma^2) + g_\rho^2\Bigl({3\pi^2\over 4} 
-{257\over 36}\Bigr)\biggr]\;.
\end{equation}
In the `$\rho$ only' case, we can drop the term with $d_{1\sigma}$ in 
(\ref{eqdpr}).  
In the degenerate case $m_\sigma=m_\rho$ with $g_\sigma=g_\rho$, 
(\ref{eqdpr}) simplifies to 
\begin{equation}
d'_{11} m_\rho^2 = 1 - {g_\rho^2 m_\rho\over 48\pi\Gamma_\rho} 
+ {m_\rho^2\over 96\pi^2 \fpi^2}\log{m_\rho^2\over \mu^2}  
-{g_\rho^2\over 16\pi^2}\Bigl(\pi^2-{173\over 18}\Bigr)\;.
\end{equation}

This procedure gives a $P$ wave phase shift with the $\rho$ mass 
and width reproducing the experimental value.  It may be thought as 
renormalizing the coupling with the `on-shell' $\rho$ width, though 
not in the sense of conventional perturbative expansion.  
Shown in Fig.~1 is the $P$ wave phase shift for three cases, $\rho$ 
only for KSRF coupling (\ref{eqKSRF}) and degenerate $\rho$--$\sigma$ for 
Veneziano (\ref{eqVeneziano}) and KSRF/Weinberg (\ref{eqWeinberg}) couplings.  
The difference in the `bare' coupling $g_\rho$ gives very slight change 
in the phase shift.  There is a small difference in the region away 
from the resonance depending whether the $\sigma$ exists or not.

We can now determine the two other subtraction constants $d'_{00}$ and 
$d'_{20}$ from the relations discussed at the end of the previous section: 
(\ref{eqdsvec}) for the $\rho$-only case, (\ref{eqdsequal}) for the 
degenerate case.  
It is then possible to calculate the two $J=0$ phase shifts using these 
parameters.

In Fig.~2(a), we show the calculated $I=J=0$ phase shift in the $\rho$ only 
scenario for three choices of $g_\rho$ (KSRF/Weinberg, Veneziano, and 
an intermediate coupling $g_\rho^2=0.45m_\rho^2/f_\pi^2$).  The experimental 
data~\cite{Protopopescu,Grayer,Alekseeva,Cason} are also plotted.  
Although the reflection of the crossed channel $\rho$ exchange gives a 
substantial effect, it can account 
only about half of the observed phase shift.   The phase shift in the 
degenerate $\rho$--$\sigma$ scenario for the same couplings is shown 
in Fig.~2(b).   The agreement with the data is reasonable.  It is rather 
difficult to determine the best value of the coupling from this data.  

The phase shift for the exotic channel $I=2$, $J=0$ is shown in Fig.~3 
with the experimental data~\cite{Walker,Colton,Cohen,Losty,Hoogland,Prukop}.  
The $\rho$ exchange [Fig.~3(a)] gives slightly larger phase shifts than the 
data.  Unlike the $I=0$ phase shift, 
the result is very sensitive to the magnitude of the coupling, 
especially for degenerate $\rho$--$\sigma$ exchanges [Fig.~3(b)].  
The intermediate coupling of $g_\rho^2\simeq  0.45m_\rho^2/\fpi^2$ 
reproduces the data quite well.

\section{Conclusions}

We have proposed a general `model-independent' framework of the $\pi\pi$ 
scattering based on chiral low-energy expansion and possible resonances 
in the $I=J=0$ and $I=J=1$ channels.  To cope with the strong interaction 
of pions, we use the $N/D$ formalism to obtain 
partial wave amplitudes which satisfy unitarity, analyticity, and approximate 
crossing symmetry.  The result is compared to the experimental phase shift 
data and we find preference for a $\sigma$ resonance with a mass similar to 
the $\rho$ meson.   Without $\sigma$, the $\rho$ exchange in the crossed 
channel can give substantial reflection in the scalar channel, but the 
effect is not large enough to explain the measured phase shift.

In this work, we have examined two clearcut cases with $\rho$ only, and 
degenerate $\rho$--$\sigma$ with the same coupling strengths.   
There is certainly some room to improve the fit if we regard the $\sigma$ 
mass and coupling as free parameters.  It is also desirable to include the 
effect of the pion mass, which we have neglected in the present study.  
These questions will be addressed in a future study.

Theoretically, $\pi\pi$ scattering is the simplest laboratory of the 
low-energy strong interaction.   Unfortunately, no new experiment has 
been done since early 1980's and the most recent result is in some 
disagreement with older data.   (We note that more recent experiments 
on the $\sigma$ meson utilize `pomeron-pomeron' scattering or $p\bar p$ 
annihilation.)  New experiments with more precision are clearly desirable.  
Systematic uncertainties may also be reduced.  In fact, 
the existing data involve some extrapolation because they are extracted 
from the reaction $\pi N \to \pi\pi N$.  It would be much more welcome 
if direct beam-beam $\pi\pi$ experiment can be done.

\section*{Acknowledgments}

We would like to thank M.\ and T.~Ishida for providing the numerical 
$\pi\pi$ phase shift data, and M.~Chanowitz and M.~Tanabashi 
for stimulating discussion.   K.~H. is supported in part by the 
Grant-in-Aid for Scientific Research (Nos.~08640343 and 10640243) 
from the Japan Ministry of Education, Science, Sports, and Culture.

\begin{figure}
\vfil
\centerline{\psfig{figure=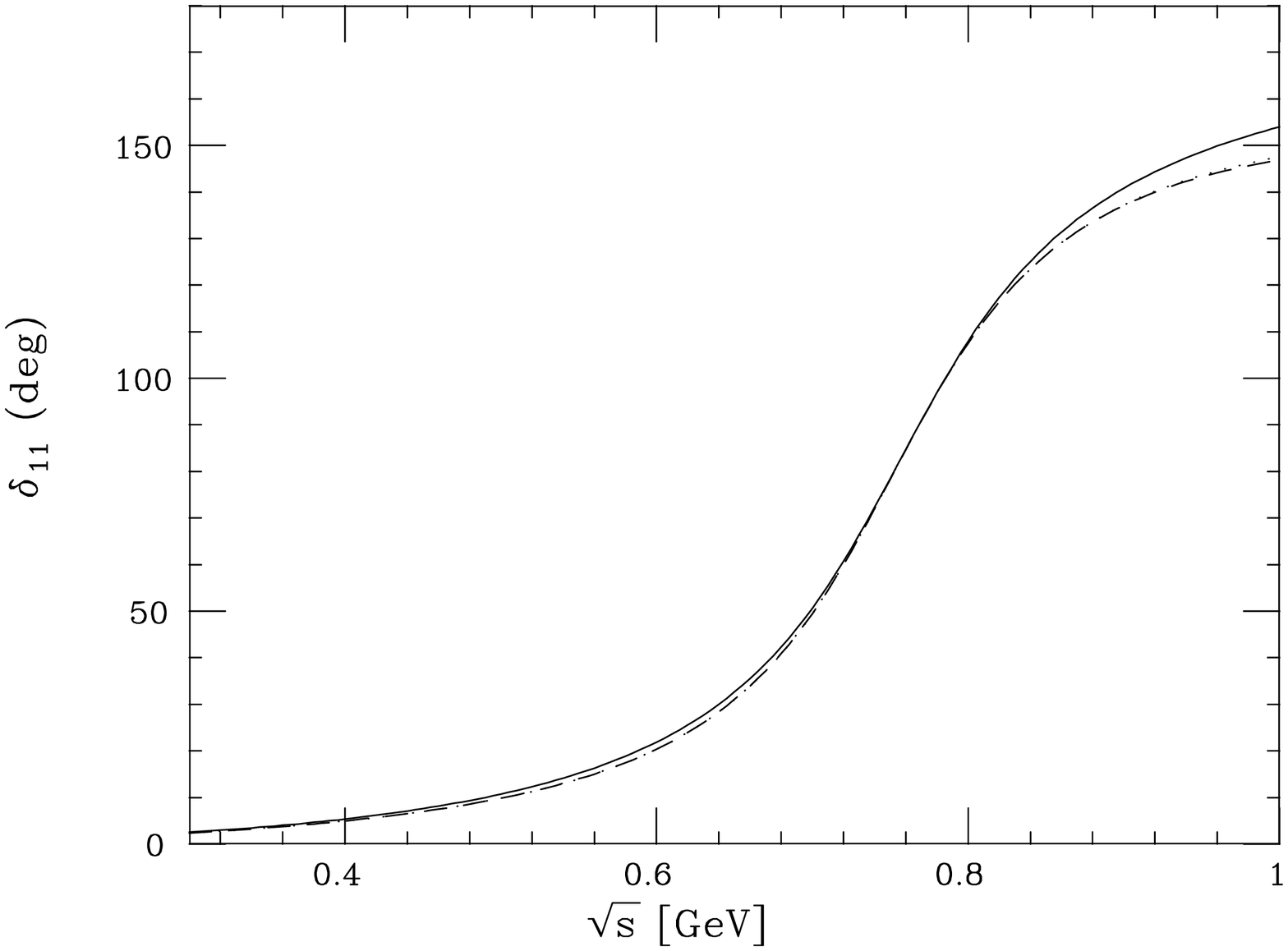,width=350pt,angle=270}}
\hspace*{10mm}
\caption{The $I=J=1$ $\pi\pi$ phase shift.  
The solid curve is for only $\rho$ exchange with the KSRF coupling, 
and the dashed (dotted) curve for degenerate $\rho$ and $\sigma$ 
with the Veneziano (KSRF/Weinberg) coupling. The latter two curves 
are almost indistinguishable.}
\label{fig1}
\vfil
\end{figure}

\eject\hrule height 0pt
\begin{figure}
\vfil
\centerline{\psfig{figure=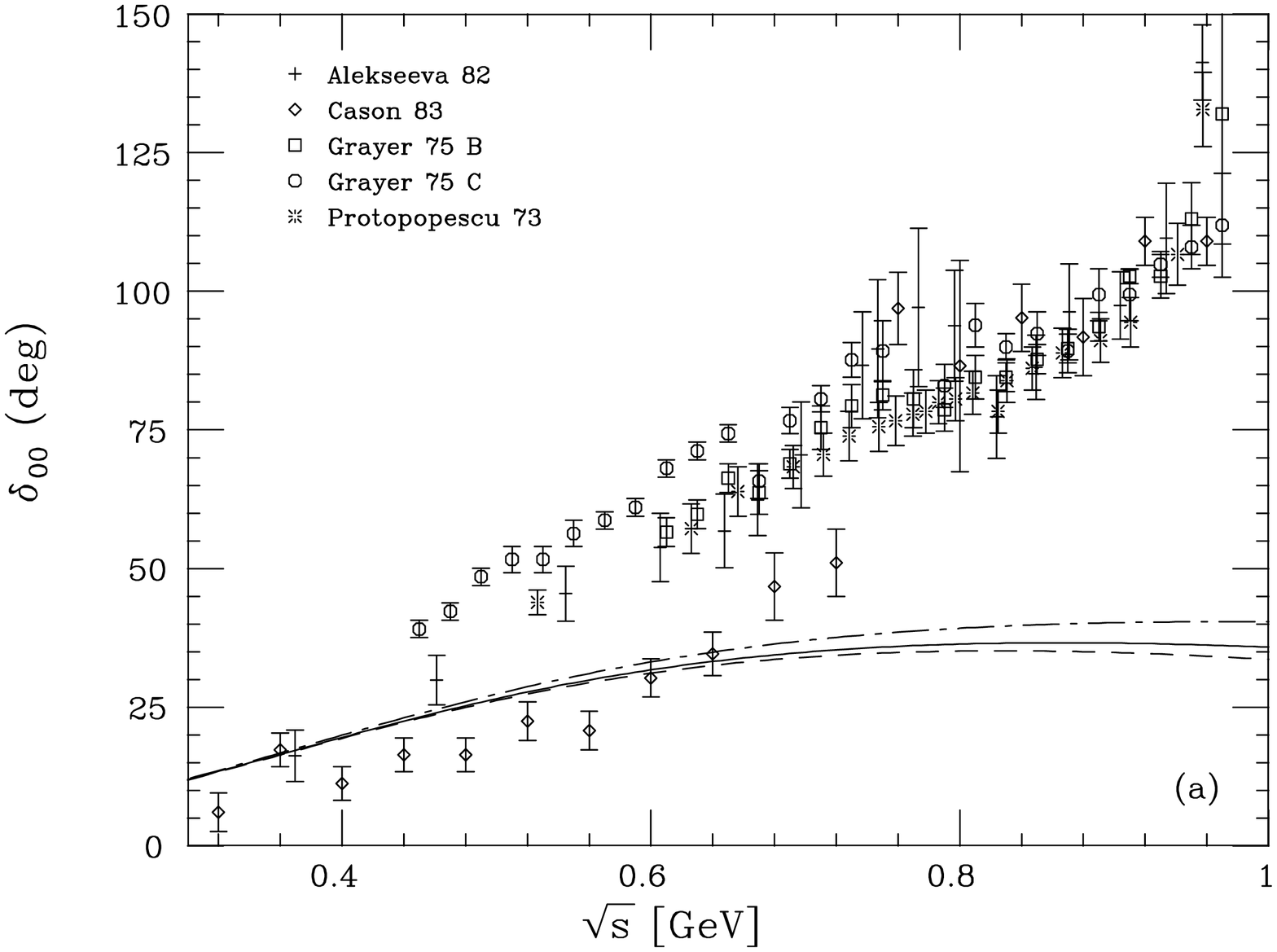,width=320pt,angle=270}}
\vspace{3mm}
\centerline{\psfig{figure=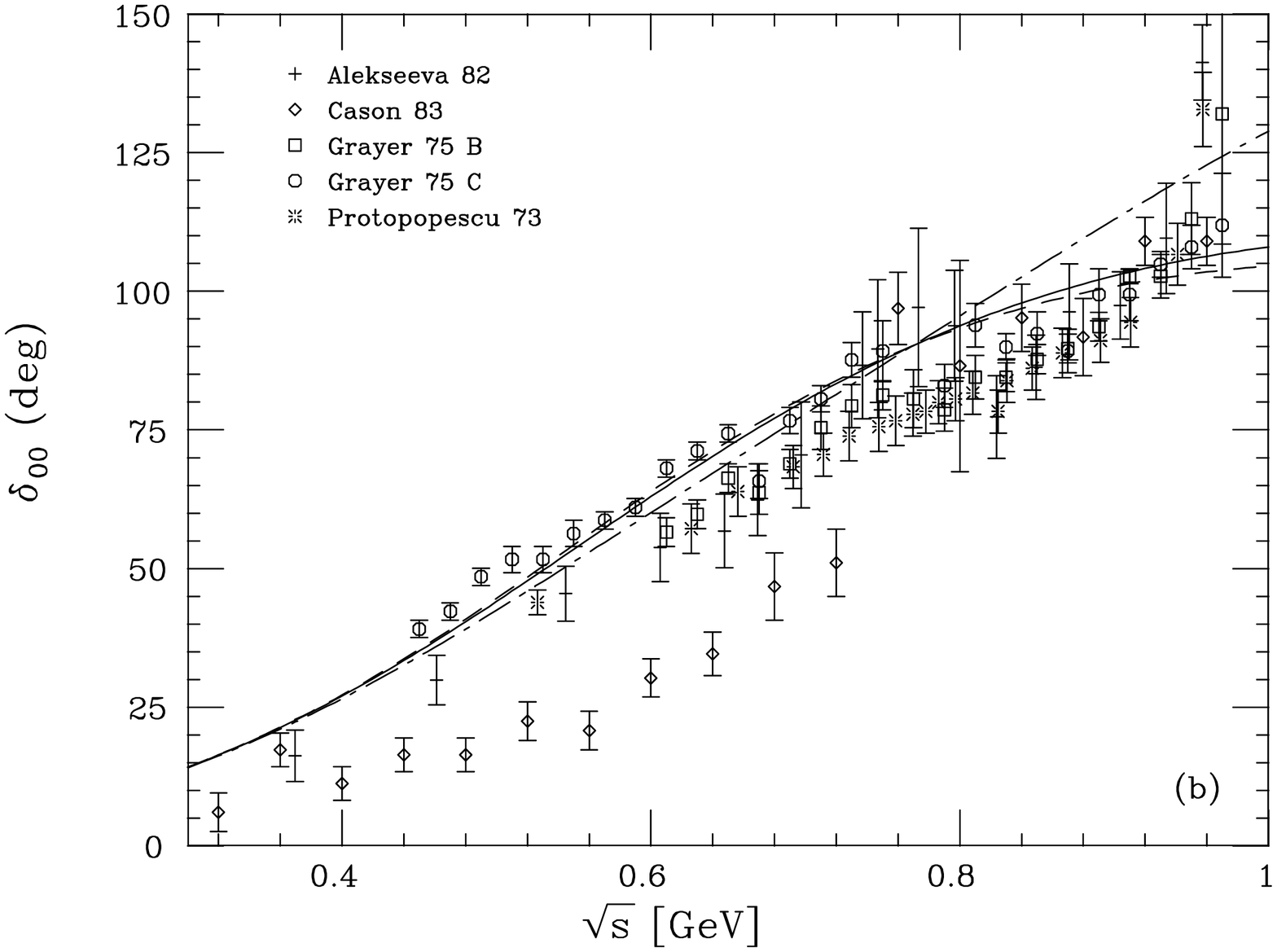,width=320pt,angle=270}}
\vspace{10mm}
\caption{The $I=J=0$ $\pi\pi$ phase shift 
with (a) $\rho$ exchange only; (b) degenerate $\sigma$ and $\rho$ exchanges 
for the KSRF/Weinberg (dashed), Veneziano (dot-dash), and intermediate 
$g_\rho^2=0.45 m_\rho^2/2\fpi^2$ (solid) couplings.  
Some experimental data are also shown.}
\label{fig2}
\vfil
\end{figure}
\eject\hrule height 0pt
\begin{figure}
\vfil
\centerline{\psfig{figure=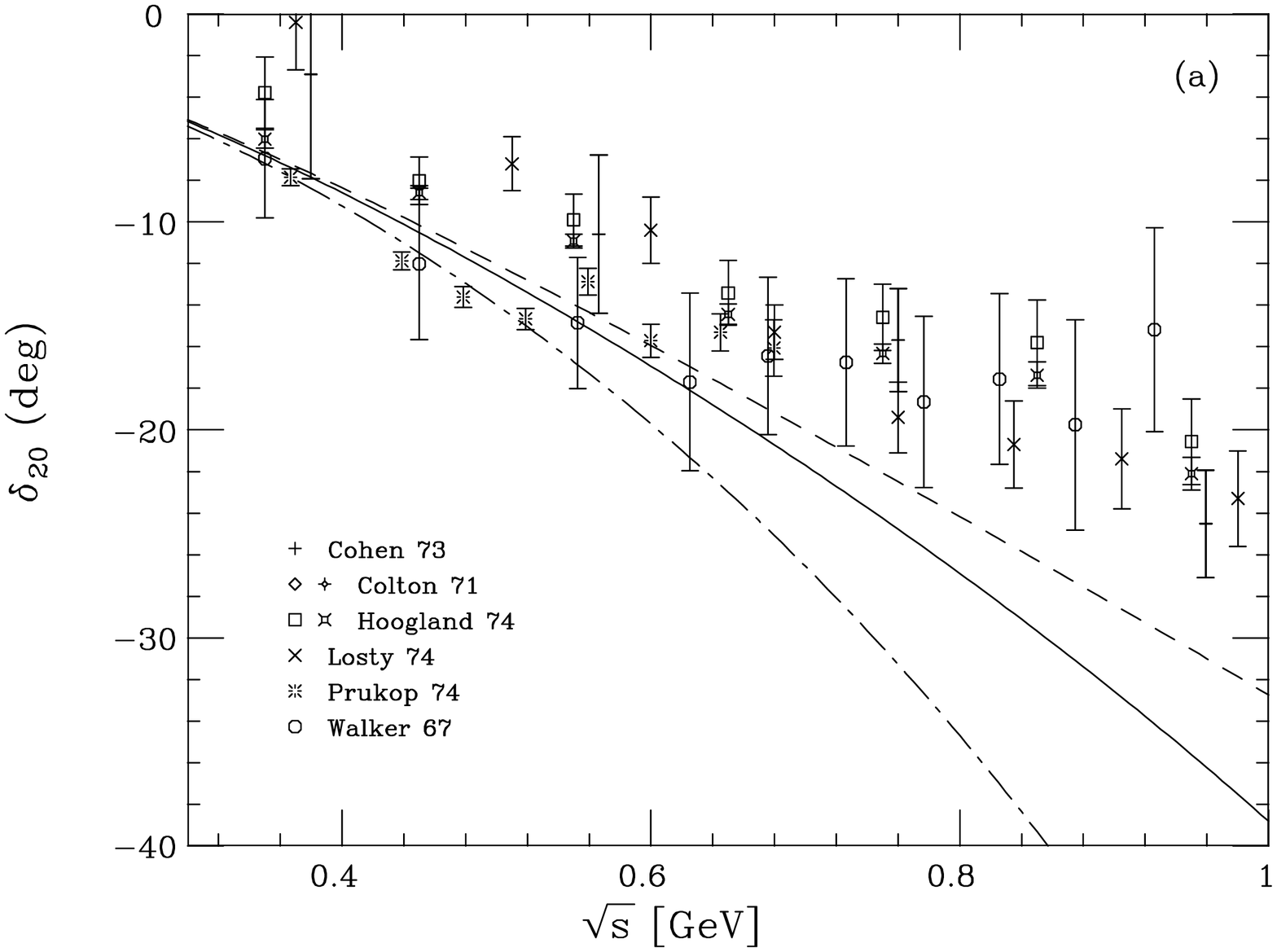,width=320pt,angle=270}}
\vspace{3mm}
\centerline{\psfig{figure=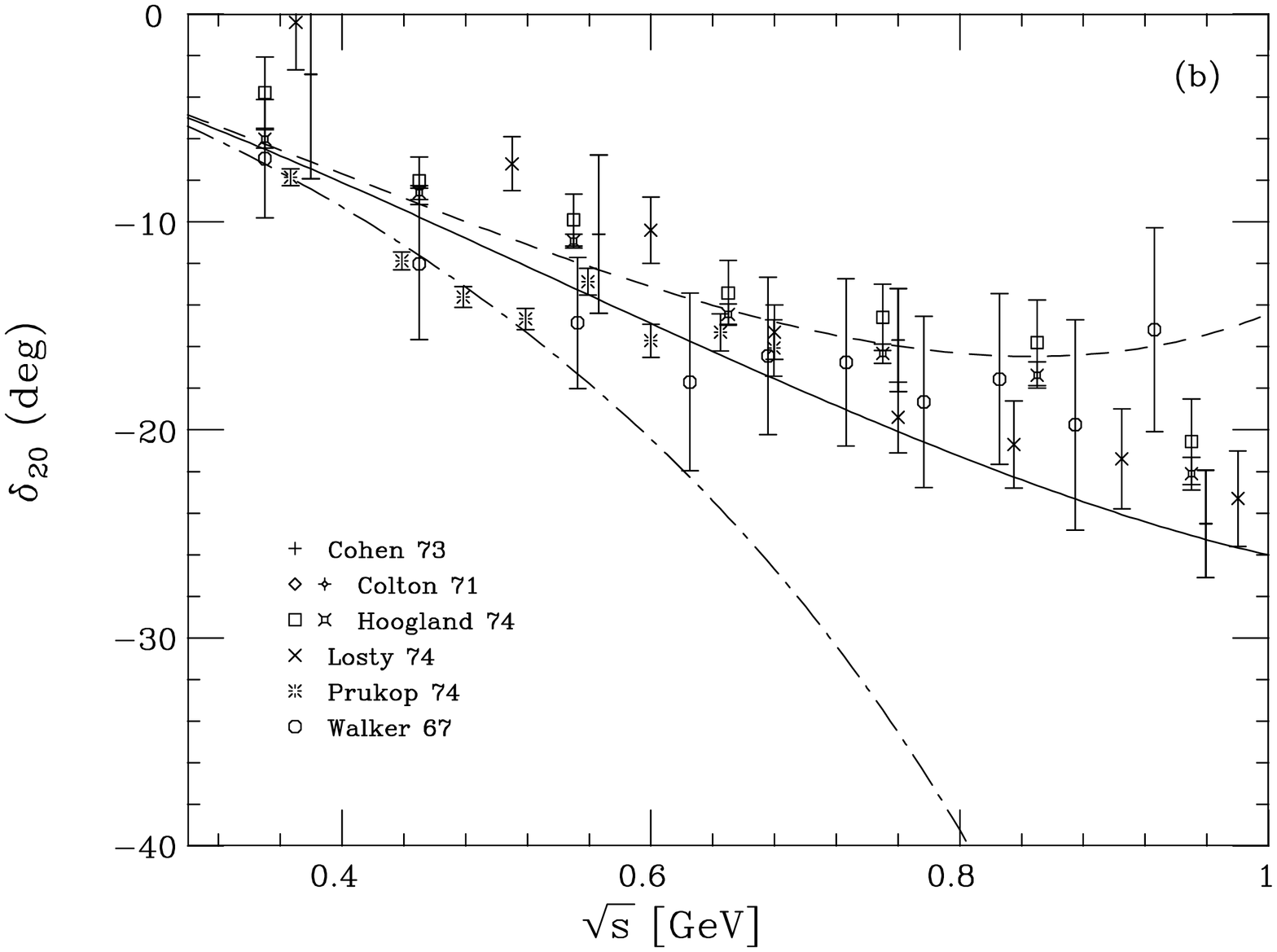,width=320pt,angle=270}}
\vspace{10mm}
\caption{The $I{=}2$, $J{=}0$ $\pi\pi$ phase shift 
with (a) $\rho$ exchange only; (b) degenerate $\sigma$ and $\rho$ exchanges 
for the KSRF/Weinberg (dashed), Veneziano (dot-dash), and intermediate 
$g_\rho^2=0.45 m_\rho^2/2\fpi^2$ (solid) couplings.  
Some experimental data are 
also shown.}
\label{fig3}
\vfil
\end{figure}
\end{document}